\documentstyle[12pt,epsfig]{article}
\newcommand{\bbar}[1]{$\overline{\mathrm{#1}}$}
\newcommand{\bbare}[1]{\overline{\mathrm{#1}}}
\newcommand{\bare}[1]{\overline{#1}}
\newcommand{\mr}[1]{{\mathrm{#1}}}

\newcommand{\reff}[1]{(\ref{#1})}
\newcommand{\be}{\begin{equation}}
\newcommand{\ee}{\end{equation}}
\newcommand{\bd}{\begin{description}}
\newcommand{\ed}{\end{description}}
\newcommand{\bmat}{\begin{displaymath}}
\newcommand{\emat}{\end{displaymath}}
\newcommand{\bit}{\begin{itemize}}
\newcommand{\eit}{\end{itemize}}
\newcommand{\ben}{\begin{enumerate}}
\newcommand{\een}{\end{enumerate}}
\begin{document}
\begin{flushright}
PRA--HEP/94--9
\end{flushright}
\vspace*{1cm}

\begin{center}
{\bf What does the CCFR  measurement of the Gross--Llewelyn Smith sum
rule tell us?} \\
\vspace*{0.3cm}
{\em Ji\v{r}\'{\i} Ch\'{y}la and Ji\v{r}\'{\i} Rame\v{s}} \\
\vspace*{0.3cm}
Institute of Physics, Academy of Sciences of the Czech Republic  \\
Na Slovance 2, Prague 8, 18040 Czech Republic \\
\vspace*{0.8cm}
{\bf Abstract}   \\
\end{center}
Recently the CCFR Collaboration reported the measurement
of the Gross--Llewellyn Smith sum rule
$\int^{1}_{0}\mr{d}F_{3}^{\nu p+\bare{\nu}p}(x,Q^2=3\;\mr{GeV}^2)=
2.50\pm 0.018(\mr{stat})$. Subsequently Kataev and Sidorov analyzed the
$Q^2$--dependence of this sum rule and pointed out a discrepancy between
the results obtained via integration of the NLO fits to
$xF_{3}(x,Q^2)$ and the purely perturbative prediction. We suggest
an explanation of this disagreement and show that the result
of the CCFR measurement of the GLS sum rule integral can be
determined from the knowledge of  the value of
$\Lambda_{\bbare{MS}}$ only.

\vspace*{1cm}
\noindent
The Gross--Llewellyn Smith sum rule provides one of the important
bridges between the QCD improved quark--parton model and the
constituent quark model. Moreover, the quantity
\be
G(Q^2)\equiv \frac{1}{2}\int^{1}_{0}\mr{d}x\left(F_{3}^{\nu p}(x,Q^2)
+F_{3}^{\overline{\nu} p}(x,Q^2)\right)
\label{GLS}
\ee
is one of the few physical quantities calculated so far up to the NNLO
of perturbative QCD \cite{Larin}
\footnote{In the following $F_{3}(x,Q^2)$ stands
for the average $\left(F_{3}^{\nu
p}(x,Q^2)+F_{3}^{\overline{\nu}p}(x,Q^2)\right)/2$.}.
In \cite{CCFR} the CCFR Collaboration reported the result
\be
G(Q^2=3\;\mr{GeV})=2.50\pm 0.018(\mr{stat.})\pm 0.078(\mr{syst.}).
\label{meas}
\ee
In \cite{KS} Kataev and Sidorov analyzed this result and addressed an
important question of the $Q^2$--dependence of this quantity. They
pointed out certain discrepancy between the predictions of perturbative
QCD for this quantity and what they call experiment. In this note
we explain the
reason for their observation and argue that in a consistent treatment
of the latter there is no place for such a discrepancy. We also comment
on the procedure used in \cite{CCFR} to evaluate \reff{GLS}. However,
as the publication \cite{CCFR} contains only a very sketchy and
incomplete description of this procedure, we frequently consulted
the unpublished thesis \cite{Leung} for details.

We start by recalling the familiar fact that in QCD the quantity
$G(Q^2)$ can be evaluated in two different ways.  First, it can be
written as a purely perturbative expansion in some renormalization
scheme RS (we use $a\equiv \alpha_{s}/\pi$ as an expansion
parameter instead of $\alpha_s$)  of the form
\be
G(Q^2)=3\left(1-
a(\mr{RS})\left(1+r_{1}(\mr{RS})a(\mr{RS})+r_{2}(\mr{RS})
a^{2}(\mr{RS})+\cdots\right)\right)
\label{g}
\ee
While $r_1$ is known already for some time from
\cite{r1}, $r_2$ has been calculated, for {\em massless quarks}
and in \bbar{MS} RS, only very recently in \cite{Larin}.
The factor 3 on the r.h.s. of the above relation is of crucial
importance. In the quark--parton model it corresponds to the fact that
there are 3 constituent (valence) quarks in the proton, but in QCD it
has much deeper meaning and is based
(see \cite{Drell} for a recent review of this subject):
on the validity of
\bit
\item equal time commutation relations of currents made
from the quark fields,
\item the asymptotic freedom.
\eit
Both of these two properties are so fundamental to QCD that we cannot
question the factor 3 in \reff{g} without abandoning the perturbative
QCD itself.  On the other hand were it possible to {\em really measure}
\reff{GLS}, the result would be, as emphasized in the original papers
\cite{GLS,Adler}, extremely important for the verification of the very
foundations of QCD.

The second theoretical route to $G(Q^2)$ exploits the nonsinglet QCD
fit to the $Q^2$ dependence of $xF_{3}(x,Q^2)$, which, beside
theoretical ingredients, requires also the specification of some initial
condition. In \cite{Leung} this initial condition at some $Q_0$
 was taken in the form
\be
xF_{3}(x,Q_{0}^2)=Ax^{\alpha}(1-x)^{\beta}+Cx^{\gamma}
\label{initial}
\ee
while Kataev and Sidorov employed a simpler form, corresponding to $C=0$
in the above expression. Fitting $A,\alpha,\beta,C,\gamma$, together
with $\Lambda_{\bbare{MS}}$, to CCFR data one can compute $F_{3}(x,Q^2)$
{\em at any x and $Q^2$} and therefrom evaluate $G(Q^2)$
at any $Q^2$!  The main observation in \cite{KS} concerns the
disagreement between these two evaluations of $G(Q^2)$. Similar
observation has in fact been made already in \cite{Leung}, but there the
reason for the appearance of such a discrepancy turns out to be
different than in \cite{KS}.

In practice \reff{GLS} is very difficult to {\em really measure}.
Because of experimental limitations the accessible range of $x$ varies
with $Q^2$ and prevents the measurement of $G(Q^2)$ separately
at each value of $Q^2$. At low $Q^2$ only very low $x$ are accessible
while for large $Q^2$ only data at large $x$ are available.  Any
experimental measurement of $G(Q^2)$ therefore inevitably involves
extrapolations to unmeasurable regions and must consequently be
considered as a combination of data and some kind of theoretical
prejudice. In \cite{CCFR,Leung} this extrapolation procedure actually
exploits the result of a nonsinglet QCD fit to $F_{3}(x,Q^2)$!  So the
experimental measurement of $G(Q^2)$ actually assumes the validity of
perturbative QCD. There is nothing illegal on this assumption, but, as
emphasized above, we have to be consistent and include in the fitting
procedure the constraint leading to the factor 3 in \reff{g} as well.
While the original determination \cite{CCFR,Leung} {\em does include}
this constraint on the QCD fit, the analysis of \cite{KS} it
{\em does not}.

Although the thesis \cite{Leung} contains many essential
details on the determination of $G(Q^2)$, it fails to
provide convincing and full description of the NLO QCD fits used in the
evaluation of $G(Q^2=3\;\mr{GeV}^2)$ at so small $Q^2$. First, there is
the question of the number $n_f$ of massless quarks used in the
evolution equations.
As these are written for a fixed number of massless quarks, one
must employ some kind of approximate procedure for crossing of
quark mass thresholds. This is important in particular for the
region of low $Q^2$, below the charmed quark threshold, which gives
nonnegligible contribution to $G(Q^2)$ but which is certainly
ineffective at $Q^2=3$ GeV$^2$. On the other hand some formulations
in \cite{Leung} indicate that the fits used in determination of
$G(Q^2=3\; \mr{GeV}^2)$ used data with $Q^2>12\; \mr{GeV}^2$
only! In this case $n_f=4$ would be a good
approximation, but then the fits {\em do not} actually fit the data
at $Q^2$ around $3\; \mr{GeV}^2$! Although our argument is independent
of the quality of these fits, this point should certainly be clarified.

The problem of crossing the quark mass thresholds is only one of the
subtle points in determining the GLS sum rule for so low $Q^2$. There
are others as well:
\bit
\item Target mass corrections. These can be easily included via
the procedure of \cite{GP}, but were ignored in extrapolation
procedure of \cite{CCFR,Leung}.
\item Higher twist terms. For $Q^2=3$ GeV$^2$ these turn out to be
{\em very important} \cite{CHK}!
\item Cabbibo mixing. Below the charmed quark threshold the $d-c$
is ineffective, thereby reducing the quantity
$G(Q^2)$ by a factor $\cos^{2}\vartheta_{c}$.
\eit
In \cite{CCFR}
the reason for evaluating the quantity $G(Q^2)$ at so low $Q^2$ was
motivated by the fact that low $x$ region gives dominating contribution
to it and 3 is just the evarage value of $Q^2$ for the lowest $x$ value
measured in the CCFR experiment. However, looking at the data in this
bin (see Fig.4 below) it is clear that these data are obviously
incompatible with the pure QCD evolution used in the extrapolation!
Whether this is due to experimental problems, or physical effects not
included in the pure perturbative QCD is another matter, but it is
certainly much safer to follow the procedure employed in \cite{KS}.
There only $Q^2\ge 12.6$ GeV$^2$ was used in NLO QCD fits to CCFR data
on $F_{3}(x,Q^2)$, thereby avoiding most of the problems mentioned
above, and $G(Q^2)$ obtained by integration of these fits was compared
to the purely perturbative expression \reff{g} taken to the LO, with
the couplant $a(Q^2)$ defined to the NLO (for reasons discussed below)
and
with $n_f=4$. The main observation of \cite{KS} concerns the
discrepancy, displayed in Fig.1, between these two results and in
particular its $Q^2$--dependence. In the rest of this note we identify
the source of this discrepancy and show how to avoid it.

Recall that as a consequence of the factorization theorem in QCD
the non--singlet (NS)
structure function $F_{3}(x,Q^2)$ can be written as a convolution
\be
F_{3}(x,Q^2)=\int^1_0 \frac{\mr{d}y}{y}q_{\mr{NS}}(y,M)C_{\mr{NS}}
\left(\frac{Q}{M},\frac{x}{y},a(\mu)\right)
\label{convolution}
\ee
between the NS distribution function $q_{\mr{NS}}(x,M)$, satisfying the
evolution equation
\be
\frac{\mr{d}q_{\mr{NS}}(x,M)}{\mr{d}\ln M}=
\int^1_0 \frac{\mr{d}y}{y}q_{\mr{NS}}(y)
P_{\mr{NS}}\left(\frac{x}{y},a(M)\right)
\label{AP}
\ee
and the ``hard scattering cross--section'' $C(Q/M,z,a(\mu))$. The scales
$M,\mu$ appearing in \reff{convolution} are the so called {\em
factorization and renormalization scales} and the couplant $a(M)$
satisfies the familiar equation
\be
\frac{\mr{d}a(M)}{\mr{d}\ln M}=-ba^2\left(1+ca(M)+\cdots\right),
\label{beta}
\ee
where the first two coefficients $b,c$ are {\em unique} functions of
$n_f$. The branching function
$P_{\mr{NS}}$ as well as the hard scattering cross--section
$C_{\mr{NS}}$ admit the following perturbative expansions
\begin{eqnarray}
P_{\mr{NS}}(z,a(M)) & = & a(M)P^{(0)}_{\mr{NS}}(z)+
a^2(M)P^{(1)}_{\mr{NS}}(z)+\cdots \label{Pexpansion} \\
C_{\mr{NS}}\left(Q/M,z,a(\mu)\right) & = & \delta(z)+
a(\mu)C^{(1)}_{\mr{S}}\left(Q/M,z\right)+\cdots
\label{Cexpansion}
\end{eqnarray}
In the NLO approximation only the first two terms in each of the
expansions \reff{beta},\reff{Pexpansion},\reff{Cexpansion} are taken
into account. Turning the above expressions into the corresponding
ones for the moments defined as
\be
    F_{3}(N,Q^2)\equiv \int^{1}_{0}x^{N}F_{3}(x,Q^2)\mr{d}x
\label{mom}
\ee
and analogously for other quantities, we get an explicit expression
for  $F_{3}(N,Q^2)$ \cite{Politzer}
\be
F_{3}(N,Q^2)=q_{\mr{NS}}(N,M)C_{\mr{NS}}(Q/M,N,a(\mu))
\label{FN}
\ee
where
\be
q_{\mr{NS}}(N,M)=
A_{N}\left[\frac{a(M)}{1+ca(M)}\right]^{-d_{N}^{(0)}/b}
\left(1+ca(M)\right)^{-d_{N}^{(1)}/bc}
\label{qN}
\ee
\be
C_{\mr{NS}}\left(\frac{Q}{M},N,a(\mu)\right)=
\left[1+a(\mu)\left(d_{N}^{(0)}
\ln \frac{Q}{M}+\frac{d_{N}^{(1)}}{b}+\kappa_{N}\right)\right]
\label{CN}
\ee
and $d_{N}^{(i)}, \; i=0,1$ are the moments of the LO and NLO
branching functions $P^{(0)}_{\mr{NS}}(z)$ and $P^{(1)}_{\mr{NS}}(z)$
respectively, and the hard scattering scale
$\mu$ is in general different from the factorization scale $M$.
Recall also that while $d_{N}^{(0)}$
is unique, $d_{N}^{(1)}$ is essentially arbitrary, defining the so
called
{\em factorization convention} (FC) \cite{Politzer,ja}. Two invariants
appear in \reff{FN}: the quantity $\kappa_N$, which is independent of
both $M$ and $d_{N}^{(1)}$, but  depends for general $N$  on the choice
of the renormalization scheme of the couplant \cite{ja}; and the
constants
$A_{N}$, which are independent of anything. As argued by
Politzer \cite{Politzer}, these constants must be kept fixed
when the factorization scale $M$ or factorization convention, defined
at the NLO by $d_{N}^{(1)}$, are varied! Moreover, these constant
provide alternative way of specifying the necessary boundary condition
on the solution of the evolution equation. Note, however, that fixing
$A_{N}$ does not correspond to specifying the initial condition on
$F_{3}(x,Q^2)$ at any finite $Q^2_{0}$. Rather the constants $A_{N}$
determine the asymptotic behaviour of $F_{3}(N,M^2)$ as $M\rightarrow
\infty$.

For $N=0$, corresponding to the sum rule \reff{GLS}, the situation is
particularly simple, as $d_{0}^{(0)}=0$ and in the FC employed in
both \cite{CCFR} and \cite{KS} (called {\em universal})
also $d_{0}^{(1)}=0$. Moreover
the value of $\kappa_{0}$ is unique and equal to $-1$. Consequently,
$q_{\mr{NS}}(N=0,Q^2)=A_{0}$ is $Q^2$--independent and
\reff{FN} reduces to
\be
F_{3}(0,Q^2)=G(Q^2)=A_{0}(1-a(\mu))
\label{F0}
\ee
which coincides with the purely perturbative expression \reff{g}
in the LO provided $A_{0}=3$ and the same couplant $a(\mu)$ is used in
both formulae. In the general case, when $\mu\neq M$, the couplant
$a(\mu)$ used in \reff{Cexpansion} can be taken only to the LO
approximation, but if we identify, as is the usual practice, $\mu=M$
and use the same couplant in both \reff{Pexpansion}, \reff{Cexpansion},
$a(M)$ in \reff{F0} is in the NLO approximation.

The crucial point of this note is
the observation that for the quantity
$G(Q^2)$ the basic principles of QCD imply $A_{0}=3$! Taking this into
account in the NLO fits to the data on $F_{3}(x,Q^2)$, the integrals
over these fits {\em must coincide} with the results of purely
perturbative predictions \reff{g}! There is simply no place for the
kind of discrepancy observed in \cite{KS}. This conclusion holds {\em
independently} of the quality of the NLO QCD fit to the data. Note also
that while in the universal FC the constraint $A_{0}=3$ is equivalent to
the condition
\be
\int^1_0 \mr{d}xq_{\mr{NS}}(x,Q^2)=3
\label{qint}
\ee
for all $Q^2$,
in general FC this is {\em not the case} and \reff{qint} may depend on
$Q^2$.

The  discrepancy observed in \cite{KS} is thus a direct consequence of
the fact that
 in their NLO fits Kataev and Sidorov {\em did not impose} the
constraint $A_{0}=3$. The technique employed in \cite{KS} to write down
the QCD predictions for the $Q^2$ evolution of $F_{3}(x,Q^2)$ is based
on the idea \cite{Show,my,oni} of expanding the convolution
\reff{convolution} in the set of Jacobi polynomials
$\Theta^{\alpha\beta}_{k}(x)$, orthogonal on the interval (0,1) with
the weight $x^{\alpha}(1-x)^{\beta}$
\be
F_{3}(x,Q^2)=x^{\alpha}(1-x)^{\beta}\sum^{\infty}_{k=0}
\Theta^{\alpha\beta}_{k}(x)a^{\alpha\beta}_{k}(Q^2)
\label{Jacexp}
\ee
where the ``Jacobi moments'' $a^{\alpha\beta}_{k}(Q^2)$ are given as
linear combinations of conventional moments $F_{3}(j,Q^2)$ \reff{FN}:
\be
a^{\alpha\beta}_{k}(Q^2)\equiv
\sum^{k}_{j=0}c^{\alpha\beta}_{kj}F_{3}(j,Q^2)
\label{Jacmom}
\ee
In \cite{KS} the boundary condition, necessary to specify such a
convolution, is taken in the form of the initial condition
\reff{initial} with $C=0$. From the fit of $F_{3}(x,Q^2)$ to the CCFR
data at all $Q^2>12$ GeV$^2$ its authors
then determine the values of the parameters
$A,\alpha,\beta, \Lambda_{\bbare{MS}}$, describing $F_{3}(x,Q^2_{0})$,
and evaluate $G(Q^2)$ at $Q_{0}^2$. Repeating this procedure with
$Q_{0}^2$ corresponding to each of the 14 values of $Q^2$ at which the
data of \cite{CCFR} exist, yields the points with error bars in Fig.~1.
Their comparison with the purely perturbative LO results
(continuous curves) reveals the discrepancy mentioned above.
In principle even a single fit is sufficient to evaluate $G(Q^2)$ at
all $Q^2$, but to integrate the expansion \reff{Jacexp} multiplied by
$1/x$ down to $x=0$ is potentially dangerous because of the oscillatory
character of the Jacobi polynomials. To integrate the initial condition
\reff{initial} is, on the other hand, a straightforward and safe matter.
It must, however, be kept in mind that athough all these 14 fits to
$F_{3}(x,Q^2)$ are equally legal NLO approximations, they lead in
general to numerically {\em different} results at all $Q^2$.

To prove our claim we have repeated the analysis of \cite{KS},
evaluating for each initial $Q_{0}^2$ the value of the constant
$A_{0}$. We have used the same method of Jacobi polynomials, described
above, as in \cite{KS}. Technical aspects of our calculations (number
ot terms taken into account in the expansion \reff{Jacexp}, accuracy
achieved etc.) can be found in \cite{my}.
For the class of conditions \reff{initial} $A_{0}$ can easily be
expressed as a function of $A,C,\alpha,\beta,\gamma$
\be
A_{0}=\left[
A\frac{\Gamma(\alpha)\Gamma(\beta+1)}{\Gamma(\alpha+\beta+1)}
+\frac{C}{\gamma}\right]\frac{1}{1-a(Q_{0})}
\label{A0}
\ee
setting $\mu=Q_{0}$ in \reff{F0}. Turned around the condition
$A_{0}=3$ implies a constraint between the mentioned parameters, for
instance
\be
A=\left[A_{0}(1-a(Q_{0}))-\frac{C}{\gamma}\right]
\frac{\Gamma(\alpha+\beta+1)}{\Gamma(\alpha)\Gamma(\beta+1)}
\label{A}
\ee
In Fig. 2 we plot
        $r(Q^2_{0})\equiv A_{0}(Q_{0})/3$, as given in \reff{A0}, with
$A,\alpha,\beta$ determined from fits of \reff{Jacexp} to CCFR data,
using the parametrization \reff{initial} with $C=0$, (i.e. the one used
in \cite{KS}), as a function of $Q_{0}$. We have repeated this exercise
also for three indicated cuts on $Q^2$ used in the fits. For all of them
the same pattern of the dependence $A_{0}(Q_{0})$ emerges: for low
$Q_{0}^2$
$A_{0}$ is below 1, while for large $Q_{0}^2$ it rises above $1$.
Note that $r(Q^2_0)$ is equal to the ratio of the two results
in Fig. 1. For all chosen samples there is thus a value of $Q_{0}^2$ for
which the {\em unconstrained} fit is consistent with the condition
$A_{0}=3$. This point depends somewhat on the $Q^2$ cut, but lies
roughly around $100$ GeV$^2$. The corresponding values of
$\Lambda_{\bbare{MS}}$ and $\chi^2/d.f.$ are plotted as functions of
$Q^2_0$ in Fig.3.

In the next step we {\em imposed} (for the simplest case $C=0$) on our
fits the constraint \reff{A}, guaranteeing thus $A_{0}=3$. The dashed
curves
in Fig.3 describe the results of these constrained fits. From Figs. 2,3
we conclude that
\ben
\item  the value of extracted $\Lambda_{\bbare{MS}}$ is only weakly
dependent on the initial $Q^2_0$, but
\item  depends sensitively on the $Q^2_{\mr{cut}}$
\item  for unconstrained fits $\chi^2/d.f.$ is an {\em increasing}
 function of $Q^2_0$, which
\item  {\em decreases} with increasing $Q^2_{\mr{cut}}$.
\item  As expected, the constrained fits yield essentially the
same values of $\Lambda_{\bbare{MS}}$ and $\chi^2/d.f.$ for those
$Q^2_0$ for which $r(Q^2_0)=1$. For lower, as well as higher,
values of $Q^2_0$ the fits have substantially higher $\chi^2/d.f.$ and yield
dramatically different $\Lambda_{\bbare{MS}}$.
\een
The last point demonstrates the importance of a proper choice of the
initial $Q^2_0$ {\em if an unconstrained fit} is performed and its
correlation with the assumed form of the initial condition
\reff{initial} and the value of $Q^2_{\mr{cut}}$ parameter.
To further illustrate the difference between the constrained and
unconstrained fits and indicate the importance of the choice of
$Q^2_0$, we compare in Fig. 4 their results to CCFR data for the
sample with $Q^2_{\mr{cut}}=12$ GeV$^2$. The values of extracted
the values of the extracted $\Lambda_{\bbare{MS}}$ (see also Fig. 3)
between $260-270$ MeV are in reasonable agreement
with the result $\Lambda_{\bbare{MS}}=237\pm 36$ MeV, obtained in
\cite{Leung}.

In principle we could use more general form of the initial condition
than in \reff{initial}, adding further parameters and looking for such
forms which would exhibit flatter dependence $r(Q^2_0)$ than in Fig. 2.
In our own investigations we have taken slightly different form of
the additional term (to the conventional $Ax^{\alpha}(1-x)^{\beta}$) in
the initial condition \reff{initial} and namely
\be
F_{3}(x,Q_{0}^2)=Ax^{\alpha}(1-x)^{\beta}(1+\gamma x)
\label{altini}
\ee
The corresponding results are not significantly different and we
therefore show only some typical examples in Fig. 3a. The basic feature
of these results is, as expected, lower $\chi^2/d.f.$ of the fits.
More detailed
investigation of this question certainly makes sense but would take
us beyond the immediate subject of this note.

Summarizing the preceding considerations we claim that provided
\bit
\item the NS fits are performed {\em with the constraint}
 $A_{0}=3$, i.e. for instance \reff{A}
\item and the LO purely perturbative approximation to \reff{g} uses
the {\em same} NLO couplant $a(M)$ as the NLO fits to $F_{3}(x,Q^2)$
\eit
these two ways of evaluating the GLS sum rule $G(Q^2)$ {\em must give
exactly the same result}. The discrepancy noted in \cite{KS}
is entirely due to the fact that in their fits Kataev and Sidorov
{\em do not impose} the constraint $A_{0}=3$.

Finally, let us briefly return to the original determination of the
GLS sum rule in \cite{Leung}, which contains basically the same
observation as in \cite{KS}. There is, however, a subtle
difference between the procedures employed in these two papers.
In \cite{Leung} the value $G(Q^2=3\;\mr{GeV}^2)=2.50 \pm 0.018$ is
obtained by extrapolating the NLO fits to $Q^2=3$ GeV$^2$,
fitting this extrapolation to the form \reff{initial}
with $C=0$ and integrating the resulting analytical formula.
Using the LO term of \reff{g}, but with LO couplant $a(M)$ only,
they get, on the other hand, $2.65\pm 0.03$. They speculate that
higher order terms and/or higher twist effects are, perhaps,
responsible for this marked difference. In fact the reason is much
simpler and again related to the constraint $A_{0}=3$.
As already emphasized, in the FC used in \cite{Leung} this constraint
is equivalent to the condition
\be
\int^1_0 q_{\mr{NS}}(x,Q^2)\mr{d}x=3
\label{Aq}
\ee
which is explicitly imposed on their fits. Had they integrated the
extrapolated
form of their fits at $Q^2=3\; \mr{GeV}^2$, they would get {\em exactly}
the result \reff{F0} with $a^{\mr{NLO}}(Q)$, the NLO couplant. Had they
furthermore used the purely perturbative expression \reff{g} to LO, but
with $a^{\mr{NLO}}(Q)$, their two results would coincide.
However, by fitting the extrapolated form of $F_{3}(x,Q^2)$ to
\reff{initial} (with $C=0$) there is no more any guarantee that the
resulting fit will satisfy the constraint $A_{0}=3$. And indeed, it
{\em does not}! Substituting the values of $A,\alpha,\beta$, published in
\cite{Leung,CCFR} to \reff{A0} we find $A_{0}=2.73$! This, together with
the fact that the purely perturbative result with the NLO couplant
equals (for the fitted value of $\Lambda_{\bbare{MS}}=237\pm 36\;
\mr{MeV}$) $2.71\pm 0.03$, explains quantitatively most of the effect
($(2.73/3)\times 2.71=2.46$).
Note that the value $\Lambda_{\bbare{MS}}=210\pm 28\; \mr{MeV}$
mentioned in \cite{CCFR,CCFR2} comes from fit to a {\em combined} $F_2$
(at large $x$) and $F_3$ data.

\newpage

\newpage
\parindent 0.01cm
{\large \bf Figure captions}

\vspace*{0.7cm}

\begin{description}
\item {\bf Fig.~1:} The quantity $G(Q^2)$ as obtained by purely
perturbative expansion \reff{g} (solid and dashed lines)
compared with its evaluation via integration of non--singlet fits to
$F_3(x,Q^2)$ (points with error bars). Dashed lines correspond to
$1 \sigma$ error of $\Lambda_{\bbare{MS}}$.
Taken from the ref. \cite{KS}.
\item {\bf Fig.~2:} $Q^2_0$ dependence of the quantity $r(Q^2_0)=A_0/3$
extracted from nonsinglet fits to $F_3(x,Q^2)$ for three different
cuts on $Q^2$ and two options on $\gamma$ in \reff{altini}.
Typical errors (not shown) of $r(Q^2_0)$ are $0.03$.
\item {\bf Fig.~3:} $\Lambda_{\bbare{MS}}$ and $\chi^2/d.f.$
corresponding to nonsinglet fits with variable initial $Q^2_0$ for the
same three classes of fits as in Fig. 2 and $\gamma=0$. In a) also the
results corresponding to variable $\gamma$ are displayed.
Errors (not displayed) of $\Lambda_{\bbare{MS}}$ are typically
$40\;\mr{MeV}$.
\item {\bf
Fig.~4:} The results of our fits to CCFR data on $F_3(x,Q^2)$ for
two values of initial $Q^2_0=10,100\;\mr{GeV}^2$, with or without the
constraint $A_0=3$. Only data at
$Q^2>12\;\mr{GeV}^2$ are included in the fits.
\end{description}
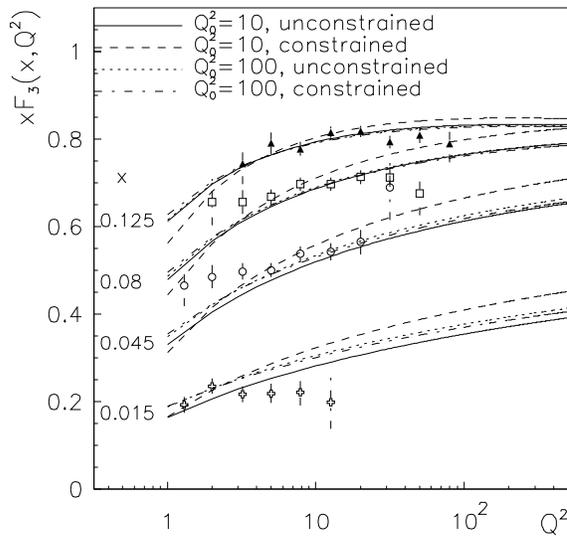
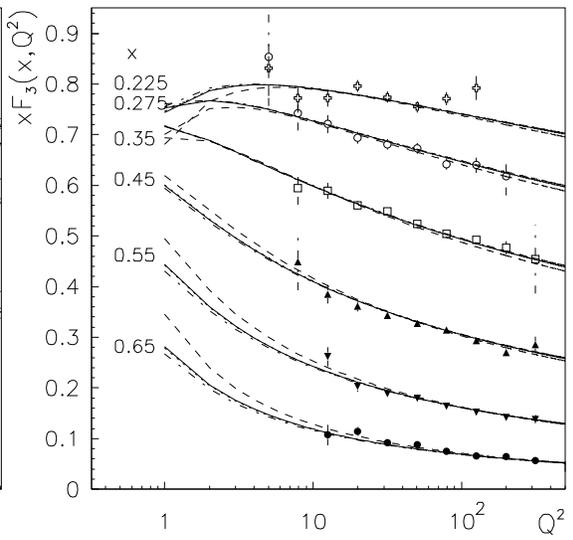
\end{document}